\documentclass[journal=jacsat,manuscript=article]{achemso}

\usepackage[version=3]{mhchem} 
\usepackage[T1]{fontenc}       
\usepackage{graphicx}		   
\usepackage{color}

\author{Vladimir R. Tuz}
\affiliation[jilin]
{International Center of Future Science, State Key Laboratory on Integrated Optoelectronics, College of Electronic Science and Engineering, Jilin University, 2699 Qianjin Street, Changchun 130012, China}
\alsoaffiliation[rian]
{Institute of Radio Astronomy of
National Academy of Sciences of Ukraine, 4, Mystetstv Street, Kharkiv 61002, Ukraine}
\email{tvr@rian.kharkov.ua}
\author{Vyacheslav~V.~Khardikov}
\alsoaffiliation[jilin]
{International Center of Future Science, State Key Laboratory on Integrated Optoelectronics, College of Electronic Science and Engineering, Jilin University, 2699 Qianjin Street, Changchun 130012, China}
\affiliation[school]
{School of Radio Physics, V. N. Karazin Kharkiv National University, 4, Svobody Square, Kharkiv 61022, Ukraine}
\alsoaffiliation[rian]
{Institute of Radio Astronomy of
National Academy of Sciences of Ukraine, 4, Mystetstv Street, Kharkiv 61002, Ukraine}
\email{khav77@gmail.com}
\author{Yuri S. Kivshar}
\affiliation[nonlinear]
{Nonlinear Physics Centre, Australian National University, Canberra ACT 2601,
Australia}
\email{ysk@internode.on.net}

\title[All-dielectric]
  {All-dielectric resonant metasurfaces with a~strong toroidal response}


\begin{document}


\begin{abstract}
We demonstrate how to create all-dielectric metasurfaces with a strong toroidal response by arranging two types of nanodisks into asymmetric quadrumer clusters. We demonstrate that a strong axial toroidal response of the metasurface is related to conditions of the trapped (dark) mode that is excited due the symmetry breaking in the cluster. We study the correlation between the toroidal response and asymmetry in the metasurface and nanocluster geometries, which appears from the different diameters of nanodisks or notches introduced into the nanodisks.
\end{abstract}

\newpage

Toroidal multipoles appear as fundamental electromagnetic excitations different from the familiar electric and magnetic multipoles~\cite{Zheludev_Science_2010}. It is known that electric multipoles result from positive and negative charges positioned over a distance, whereas magnetic multipoles are produced by electric currents circulating on a contour. In contrast, the simple member of the toroidal multipoles, a magnetic (\textit{polar}) toroidal dipole, is created by poloidal electric currents flowing on a surface of a torus along its meridians. Such a current flow can be represented as a set of magnetic dipoles arranged head-to-tail to form a closed ring. A dual counterpart of the polar toroidal dipole is an electric (\textit{axial}) toroidal dipole, which is composed by a ring of the electric dipolar configurations \cite{Dubovik_JETP_1986, Thorner_PhysRevB_2014}. Mathematically speaking, both electric and magnetic multipoles appear from the Taylor expansions (known as the multipole expansions) of the electromagnetic potentials and sources, whereas toroidal multipoles are obtained by the decomposition of the moment tensors~\cite{Zheludev_NewJPhys_2007, Savinov_PhysRevB_2014}. In this decomposition a magnetic toroidal dipole moment is represented by a time-odd polar vector, which changes sign under both time and space inversions. In the electric toroidal dipole moment these inversions are symmetric.

It is revealed that both types of toroidal multipoles can exist in natural media. In particular, a \textit{static} toroidal moment appears in several ferroelectric systems \cite{Naumov_Nature_2004, Spaldin_JPhysCondensMatter_2008}, biological and chemical macromolecules \cite{Ceulemans_PhysRevLett_1998}, and glasses \cite{Yamaguchi_NatureCommun_2013}. Its nature is of great interest in solid state physics because it is expected that toroidal topology can open up possibilities for a new kind of magnetoelectric phenomena with their prospect in applications for data storage and sensing \cite{Talebi_nanoph_2017}. Unfortunately, the toroidal response in natural media typically is very weak and hardly measurable since it is often masked by much stronger electric and magnetic multipoles.

Importantly, the toroidal effects can be enhanced significantly in artificial materials (metamaterials) configured on demand \cite{Zheludev_NatMater_2012, Zheludev_Science_2015}. In particular, an appearance of a \textit{dynamic} toroidal moment has been demonstrated for the first time in the microwave band in the metamaterial constructed from a periodical arrangement of clusters (meta-molecules) of four metallic split wire loops \cite{Zheludev_Science_2010}. The conductive currents excited by an incident plane electromagnetic wave in the arcs of such a meta-molecule  resemble the flow similar to that existing in toroidal wire coils \cite{Papasimakis_PhysRevLett_2009}. Such current configuration provides suppressing of the magnetic and electric moments, against which a toroidal response arises. This approach is further developed in other designs based on split-ring and double-ring resonators (or their metallic counterparts) for operating in microwave \cite{Dong_OptExpress_2012, Guo_EurPhysJB_2012, Fan_PhysRevB_2013, zheludev_PhysRevB_2016, Stenishchev_SciRep_2017}, infrared \cite{Dong_ApplPhysLett_2012, Dong_PhysRevB_2013, Gupta_ADMA_2016}, and visible \cite{Ogut_NanoLett_2012, Huang_OptExpress_2012, Huang_SciRep_2013} parts of spectrum.

Nevertheless, the implementation of toroidal response based on metamaterials with metallic inclusions operating in the infrared and visible spectra runs into serious obstacles. In fact, metallic particles in a meta-molecule capable to support a toroidal response usually have a very intricate shape, and, thus, there is an apparent technological problem of their manufacturing on the nanometer scale. Another severe problem is Ohmic losses and, thus, huge energy dissipation in metallic components of the meta-molecule which inevitably appears at the optical frequencies. In order to resolve these issues, we should employ all-dielectric metamaterials~\cite{Decker_AdvOptMat_2015, jahani_NatNano_2016, Kuznetsov_Science_2016, Kruk_ACSPhotonics_2017}.

All-dielectric metamaterials consist of specially arranged dielectric particles or clusters based on them. Each single particle ("meta-atom") is made of a high-refractive-index low-loss material, and it behaves as an open volumetric dielectric resonator. The resonator supports a variety of electric and magnetic dipolar and multipolar modes (referred to Mie-type modes \cite{Zhao_MatToday_2009}) associated with strong \textit{displacement} (polarization) currents induced inside the particle by an incident electromagnetic wave, and appearance of these modes can be spectrally controlled and engineered independently. However, since the particles have a simple shape being made of a nonmagnetic material, a toroidal response can be excited in a single particle of a larger size \cite{Miroshnichenko_NatCommun_2017}, but  to excite a strong toroidal moment, we should consider a cluster (meta-molecule) of dielectric particles \cite{Basharin_PhysRevX_2015, Soukoulis_PhysRevB_2016, Stenishchev_SciRep_2017}.

Depending on the light polarization and spatial symmetry of the meta-molecule forming the metamaterial, different strategies are proposed to excite the toroidal moment. Thus, an electromagnetic wave, with either linear \cite{Gupta_ADMA_2016}, radial \cite{Bao_SciRep_2015}, or circular polarization \cite{Gui_ApplPhysLett_2014, zheludev_PhysRevB_2016}, has been used to illuminate the metamaterials normally, laterally, or obliquely (in fact, split-ring resonator's based metamaterials usually are illuminated from the lateral direction). In general, the polarization of the electric field of the incident light induces effective current loops in the structure, which generates magnetic dipoles. By carefully tailoring the structure elements considering polarization of the incident wave, a vortex of magnetic dipoles, i.e. a toroidal moment, is thus achieved \cite{Talebi_nanoph_2017}. 

\begin{figure}[ht!]
\centering\includegraphics[width=10cm]{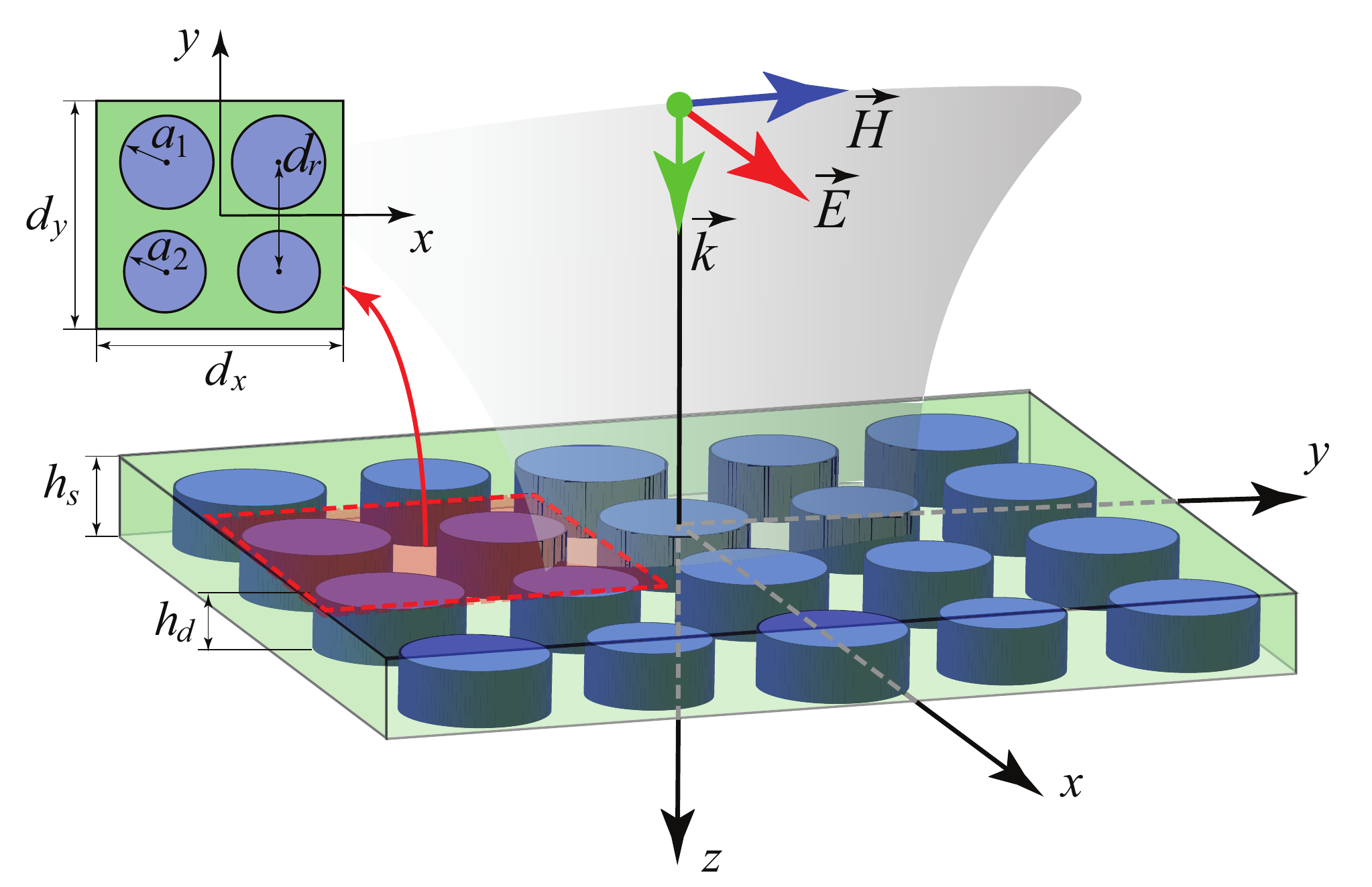}
\caption{Schematic of an all-dielectric metasurface. A meta-molecule is made of a cluster consisting of two pairs of dielectric nanodisks with different radii $a_1$ and $a_2$, so that the cluster is asymmetric with respect to the $x$-axis. lattice spacing is $d_r$. The whole structure is under an illumination of a normally incident plane wave whose electric field is directed along the $x$-axis ($x$-polarization).}
\label{fig:struct}
\end{figure}

Apart from a variety of known metamaterial's configurations supporting Mie-type multipole excitations we further distinguish a particular class of resonant metasurfaces which allow obtaining the strongest resonant response in a deeper subwavelength optically thin structure due to excitation of so-called trapped (dark) modes \cite{Zouhdi_Advances_2003, Fedotov_PhysRevLett_2007}. As known, such modes appear in metamaterials provided that their meta-molecule possesses certain structural asymmetry. In the all-dielectric meta-molecule an electromagnetic coupling between dielectric particles arises from particular antiphase oscillations of displacement currents that are weakly coupled to free space \cite{Khardikov_JOpt_2012, Khardikov2016}. 

In this Letter, we demonstrate that with a proper arrangement of dielectric particles forming a spatially \textit{asymmetric} cluster, we can realize an all-dielectric  metasurface with a strong axial toroidal response. Such a toroidal response appear from the trapped mode excitation by a normally incident linearly polarized plane electromagnetic wave. We believe this is the first example of such metasurfaces that may initiate novel experimental studies with the generation and enhancement of optical toroidal moments. 

As the first step, we demonstrate an axial toroidal response that appears in an all-dielectric metasurface whose unit cell (meta-molecule) is made of a cluster consisting of four solid nanodisks (a quadrumer). The cluster is a square ($d_x = d_y = d$), and the distance $d_r$ between the centers of disks is fixed in both $x$ and $y$ directions. All disks are made from a nonmagnetic lossless dielectric having permittivity $\varepsilon_d$ and height $h_d$. We consider that the nanodisks are grouped into pairs in which the disks differ in radius ($a_1 \ne a_2$). The pairs are then arranged within the cluster in such a way that the meta-molecule appears to be symmetric with respect to the $y$-axis, whereas with respect to the $x$-axis it is asymmetric. The parameter $\Delta$ is introduced to specify the difference in radii of the nanodisks in their pairs forming the cluster, and, thus, this parameter expresses the degree of structural asymmetry of the unit cell. Finally, the lattice of disks is buried into a dielectric host having permittivity $\varepsilon_s$ and thickness $h_s$ to form a metasurface (Fig.~\ref{fig:struct}). 


In order to excite the trapped mode, the electric filed vector $\vec E$ of the incident wave has to be oriented along the axis of the structure asymmetry. Therefore, further we define that the metasurface under study is illuminated by a normally incident plane wave $(\vec k = \{0,0,k_z\})$, whose electric field vector $(\vec E=\{E_x,0,0\})$ is directed along the $x$-axis ($x$-polarization).

We perform a numerical study of the electromagnetic response of the metasurface using the RF module as a part of the commercial COMSOL Multiphysics finite-element-based electromagnetic solver, where the Floquet-periodic boundary conditions are imposed on four sides of the unit cell to simulate the infinite 2D array of volumetric dielectric resonators\cite{comsol}. In our numerical model we suppose that disks are made of silicon ($\varepsilon_d = 13.7$), and they `hang' in the air ($\varepsilon_s = 1$). We introduced the last assumption, realizing that, in fact, if the structure is completely buried inside a dielectric host with relative permittivity $\varepsilon_s > 1$, the resonant frequencies will reduce with the factor $\sqrt{\varepsilon_s}$ \cite{munk_2000}. 

\begin{figure}[ht!]
\centering\includegraphics[width=10cm]{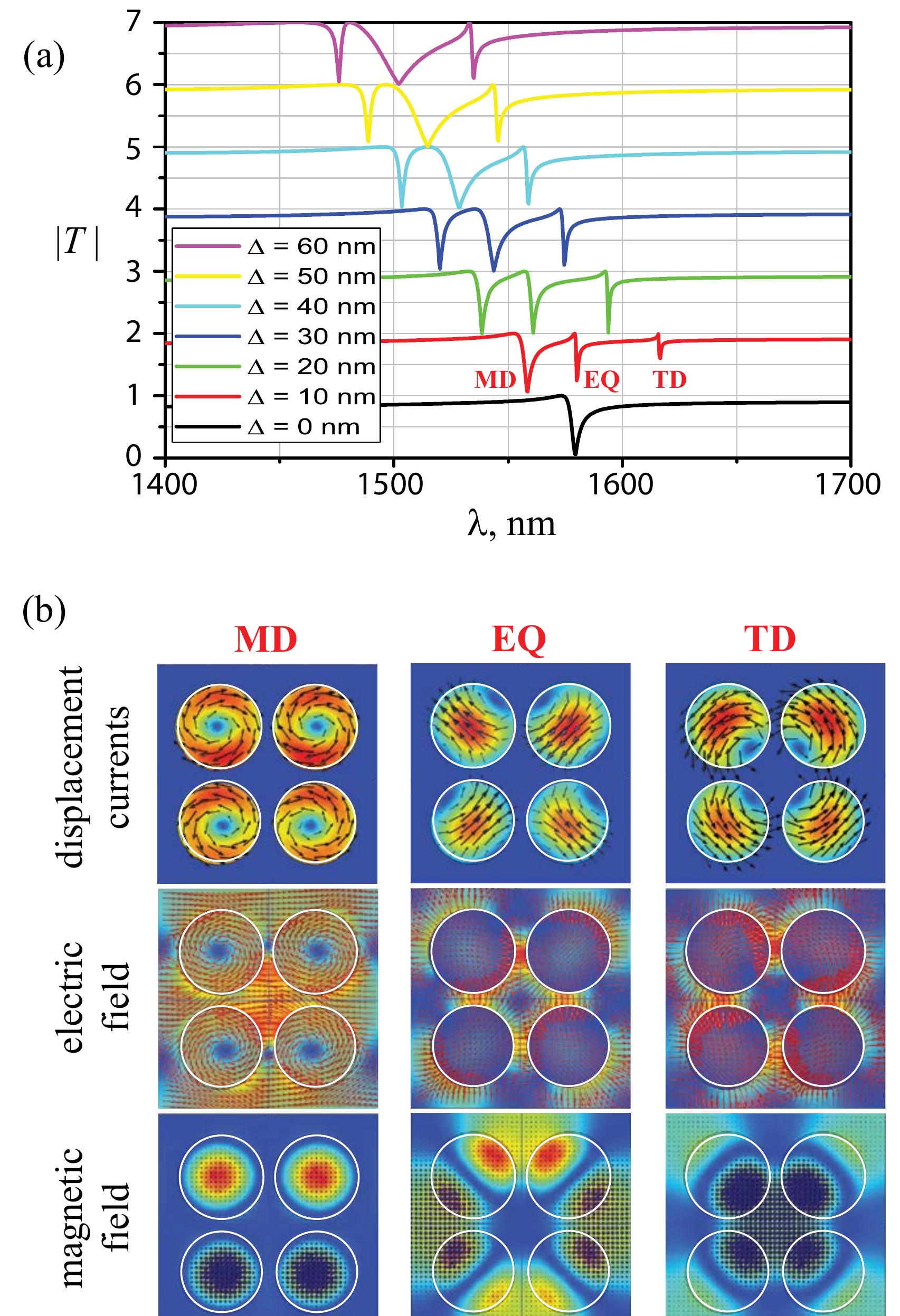}
\caption{(a) Transmission coefficient magnitude of an all-dielectric metasurface whose meta-molecule is made of an asymmetric cluster of four solid disks, and (b) cross-section patterns of displacement currents (black arrows), electric (red arrows) and magnetic (blue arrows) field distribution which are calculated within the meta-molecule at the corresponding resonant wavelength; $d= 1250$~nm, $d_r=540$~nm, $h_d=120$~nm, $a_1=240$~nm, $a_2=a_1-\Delta$.}
\label{fig:solid_disks}
\end{figure}

It is well known that in a metasurface consisting of an array of equidistantly spaced subwavelength dielectric disks each particle (meta-atom) behaves as an individual resonator sustaining a set of electric and magnetic multipole modes whose coupling to the field of the incident wave produces a strong resonant electromagnetic response of the whole metamaterial \cite{Decker_AdvOptMat_2015}. From this set of modes the first two long-wavelength resonances appear from the elementary magnetic dipole (MD) and electric dipole (ED) excitations. Their order on the wavelength scale may vary depending on the geometrical parameters of the disks. In fact, among other parameters, the resonant wavelength of the MD and ED excitations depends crucially on the disk height and radius, respectively. 

Unlike the previous situation, the unit cell of the metasurface under study is formed by a cluster which itself appears as a subwavelength meta-molecule, where four disks are arranged closely to each other. In this case an overall metamaterial response is mainly determined by the degree of electromagnetic coupling between the particles inside the cluster rather than by the individual disks. The mutual influence of neighboring clusters of the 2D lattice also has an effect. It inevitably leads to appearance of a complex collective behavior of modes which is very different from that of the elementary MD and ED excitations. 

The appearance of such a collective response is confirmed by our numerical simulations of both transmitted spectra of the metasurface and patterns of the displacement current, electric and magnetic field distributions inside the unit cell (Fig.~\ref{fig:solid_disks}). The latter are calculated at particular resonances corresponding to the certain mode excitations. For all our presented illustrations the wavelength range is chosen in such a way that only the first three long-wavelength resonances are inside the region of interest. 

Thus, in the symmetric design when all disks in both pairs of the cluster have the same radius ($\Delta = 0$), there is only one resonance in the chosen wavelength range. We distinguish it as a collective response of four longitudinal MDs. Thus, the displacement currents at the resonance  demonstrate a circular behavior twisting around the center of each disk in the $x-y$ plane producing four magnetic moments oriented along the $z$-axis (see the first column in Fig.~\ref{fig:solid_disks}b), and these moments oscillate in antiphase in disks oppositely located with respect to the $x$-axis. In fact, the displacement currents resemble behaviors of the TE$_{01}$ mode inherent to the individual cylindrical dielectric resonators\cite{Snitzer_JOptAm_61, marcatili1964hollow}. The resonance related to such a collective magnetic response acquires a sharp peak-and-trough (resonance-antiresonance) profile known as Fano one \cite{Kivshar_NatPhot_2017}. It appears since the resonance produced by MDs is overlapped with the broad electric dipole resonance which is induced by the incident electromagnetic wave on the whole meta-molecule. Moreover, this resonance is polarization-insensitive (i.e., it appears in the transmitted spectra for both $x$-polarized and $y$-polarized waves) and cannot be excited in a lattice with equidistantly spaced identical particles since the oppositely directed magnetic moments fully compensate each other in this case. 

As soon as an asymmetry in the meta-molecule is introduced ($\Delta \ne 0$), besides the resonance related to the previously discussed combination of four longitudinal MDs, two additional resonances appear in the transmitted spectra of the metasurface within the wavelength range of interest. Considering the corresponding displacement current distribution, and patterns of the electric and magnetic fields inside the cluster we relate these resonances to the electric quadrupole (EQ) and electric toroidal dipole (TD) excitations, correspondingly, from the left to right on the wavelength scale. The EQ mode evidently appears from the uncompensated electric moments, which arise in the disks with different radii. The TD mode is recognized from a peculiar closed loop of the displacement currents that penetrates all four disks in the cluster and encloses the circulating electric field lines into a torus, whereas the magnetic field lines are directed along the torus meridians. Remarkably, both modes produce the resonant spectral line having the Fano form. From the complete set of curves presented in Fig.~\ref{fig:solid_disks}a one can also conclude that as the degree of asymmetry increases all three distinguished resonances remain stable performing only a small blue-shift, while the distance between the resonances is almost unchanged. Since the asymmetric cluster geometry is inherently anisotropic, both EQ and TD resonances appear only for $x$-polarized excitation.

Remarkable, all mentioned resonances exist only in a complicated (cluster) metasurface and do not arise in a lattice with equidistantly spaced identical solid disks. We explain this fact in terms of compliance of their resonant conditions to those existing in the trapped modes. The trapped modes appear in solutions of the scattering and radiation problems, where they represents a free (eigen) oscillations with finite energy of the fluid surrounding the fixed structure. Mathematically, the trapped mode corresponds to an eigenvalue embedded in the continuous spectrum of the relevant operator \cite{Evans_MechApplMath_1993, Evans_JFluidMech_1994}. Such a combination of the discrete and continuous spectra results in the fact that the resonant characteristic of the system supporting trapped modes acquires a Fano profile. 

In particular, in the metasurface under study, for all three resonances there are branches where the corresponding displacement currents oscillate in antiphase. In the lattice having equidistantly spaced disks all moments produced by such antiphased oscillations are fully compensated (i.e., these oscillations resemble the trapped mode behaviors and are not coupled to free space). In the asymmetric structure the moments are no longer compensated, and, thus, there is a coupling of the corresponding trapped mode with the field of the incident wave, which results in the resonance arising in the electromagnetic spectra. We should note that the scattered electromagnetic far-field produced by the trapped modes is very weak, which drastically reduces the structure coupling to free space at the resonance and therefore decreases strongly the radiation losses \cite{Zouhdi_Advances_2003, Fedotov_PhysRevLett_2007}. It explains a high-$Q$ feature of the corresponding resonances. 

\begin{figure}[ht!]
\centering\includegraphics[width=10cm]{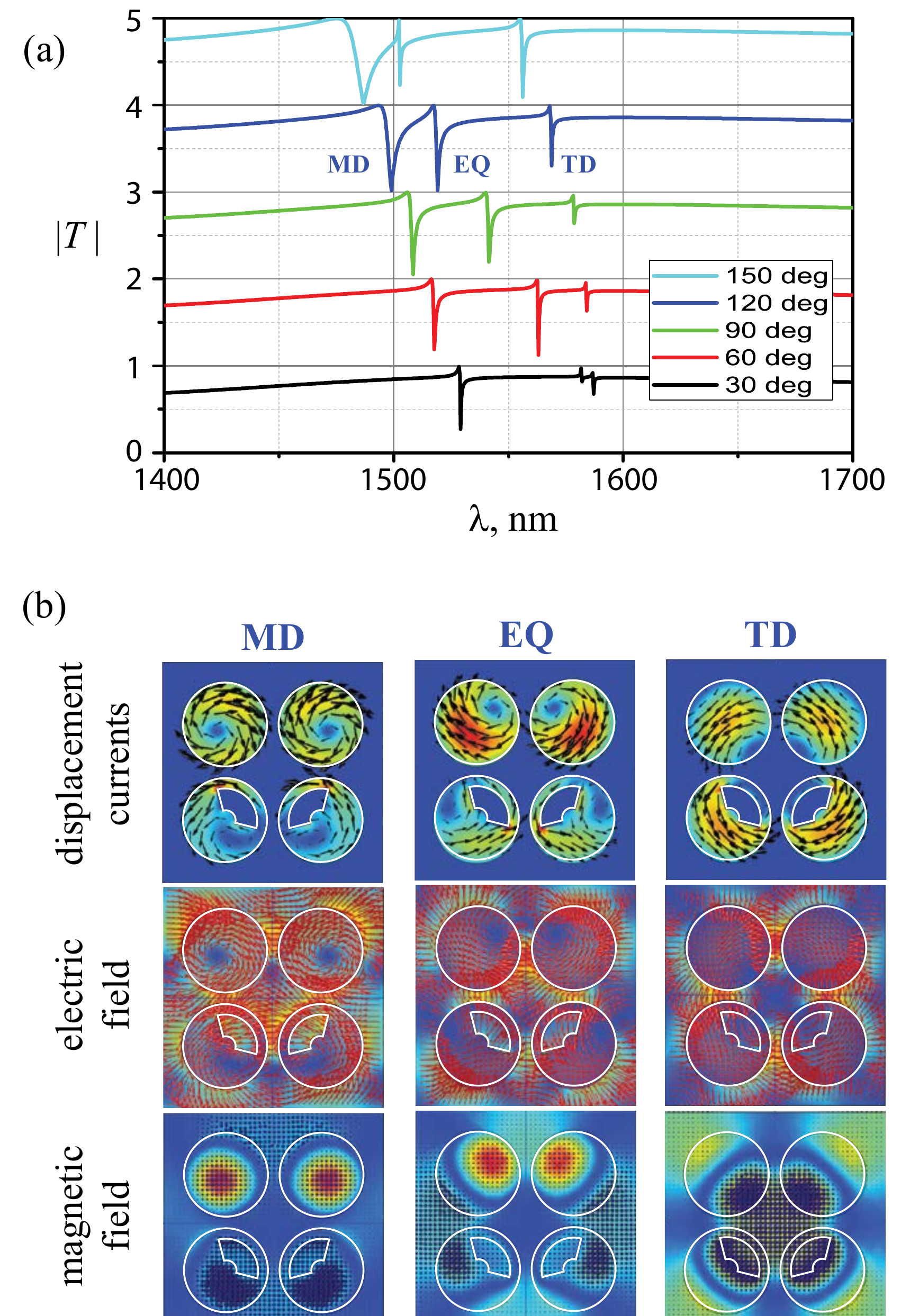}
\caption{(a) Transmission coefficient of an all-dielectric metasurface whose asymmetric meta-molecule is made of a cluster with a pair of two solid nanodisks and two nanodisks having coaxial-sector notch (smile), and (b) cross-section patterns of the displacement currents (black arrows), electric (red arrows) and magnetic (blue arrows) fields  calculated within the unit cell at the corresponding resonant wavelengths; $d= 1250$~nm, $d_r=540$~nm, $h_d=120$~nm, $a_1=240$~nm, $a_h=a_1/2$, $w=a_1/2$.}
\label{fig:smile_disks}
\end{figure}

Structural asymmetry in the metamolecule and the coupling to the trapped modes can be created in various ways. On the other hand, in order to excite the toroidal dipole mode, an arrangement of particles inside the cluster should be carefully tailored. Below, we discuss the toroidal dipole excitation for another metasurface's design, where the disks in the cluster have the same radius, while a spatial asymmetry is introduced  by a certain perturbation in the nanodisks. In particular, we propose to form a meta-molecule by combining a pair of solid disks with another pair of disks having a coaxial-sector notch (smile). We prefer this design considering that the notch has a form to affect the circulating displacement currentsthat produces a toroidal responce. 

In this geometry, $\Delta$ is the radius of the sector mid-line, $2a_h$ is the notch width, and $\alpha$ is the sector opening angle. The form of the unit cell is presented in Fig.~\ref{fig:smile_disks} from which one can conclude that in this design the cluster's asymmetry is preserved with respect to the $x$-axis, while the degree of asymmetry is defined by the sector opening angle $\alpha$. Moreover, it is obvious that the degree of asymmetry varies between two extreme angles $\alpha_{min}=0^\circ$ and $\alpha_{max}=360^\circ$ at which the meta-molecule becomes symmetric.  

\begin{figure}[ht!]
\centering\includegraphics[width=10cm]{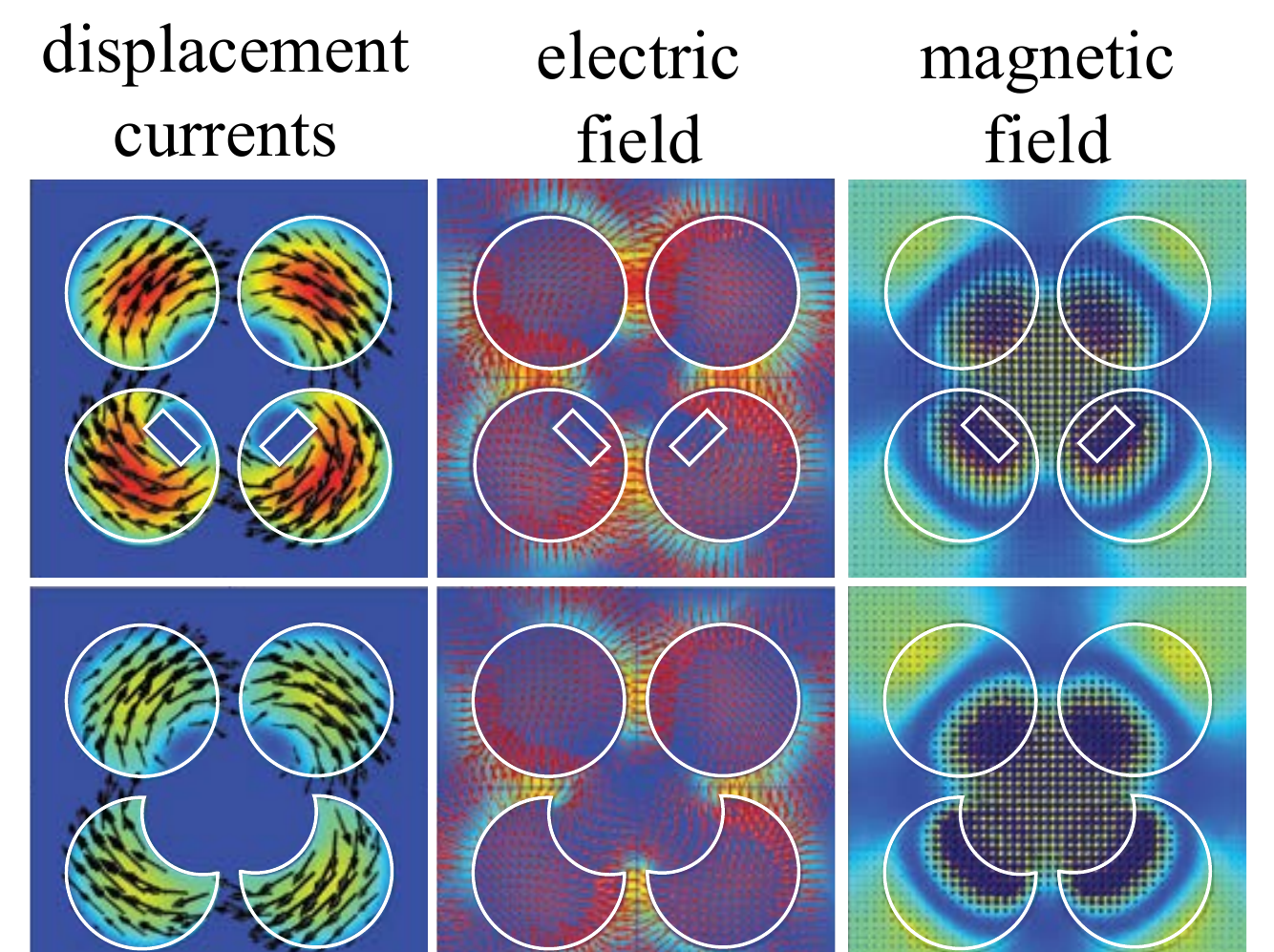}
\caption{Displacement current (black arrows), electric (red arrows) and magnetic (blue arrows) fields calculated within the unit cell of a metasurface at the corresponding wavelength of the toroidal dipole mode for two particular designs; $d= 1250$~nm, $d_r=540$~nm, $h_d=120$~nm, $a_1=240$~nm. The metamolecule is made of a cluster with either two solid nanodisks and two nanodisks with a rectangular notch (top row) or two solid nanodisks and two nanodisks with a rounded notch (bottom row); $a_h=a_1/2$, and $w=a_1/4$.}
\label{fig:modif_disks}
\end{figure}

A set of corresponding curves for the transmission coefficient calculated versus the wavelength of the incident wave is presented in Fig.~\ref{fig:smile_disks} for the proposed metasurface's design. Our simulations also yield the displacement currents distribution as well as electric and magnetic field patterns plotted inside the unit cell which allow identification of the modes associated with the calculated transmitted spectra. One can see that in the transmitted spectra there appear all previously discussed resonances, and, thus, the cluster with smile disks is suitable for the toroidal dipole excitation. Nevertheless, while the quality factor and resonant wavelength of the longitudinal MD excited by such disks are almost identical with the previously considered for the solid disks, those of the QD and TD are strongly dependent on the sector opening angle, and there is a variation of the distance between QD and TD resonances on the wavelength scale. It can be concluded that such a design gives more tuning freedoms where the notch parameters open a prospect to find an optimal solution for the meta-molecule geometry producing the maximally bright TD response. Moreover, the notch can be filled with other material possessing properties of a nonlinear or gain medium to expand the metasurface functionality \cite{tuz_PhysRevB_2010, Khardikov2016, tuz_EurPhysJApplPhys_2011, Tuz_JOSAB_2014}.

Taking into account the fact that nanodisks with complex coaxial-sector notches are difficult to fabricate for the visible spectrum, we consider  simpler designs of the nanocluster. Thus, combinations of pairs of the solid nanodisks with  commensurate nanodisks having either rectangular holes or rounded notches were additionally simulated (see Fig.~\ref{fig:modif_disks}). We find that in both these configurations the toroidal mode is still excited. Therefore, it can be concluded that the conditions for the toroidal mode excitation are determined by the presence of a structural asymmetry; it depends also on the arrangement of particles inside the cluster and
on the nature of the introduced inhomogeneity into the nanodisks, however all these parameters must be matched appropriately.

\begin{figure}[ht!]
\centering\includegraphics[width=10cm]{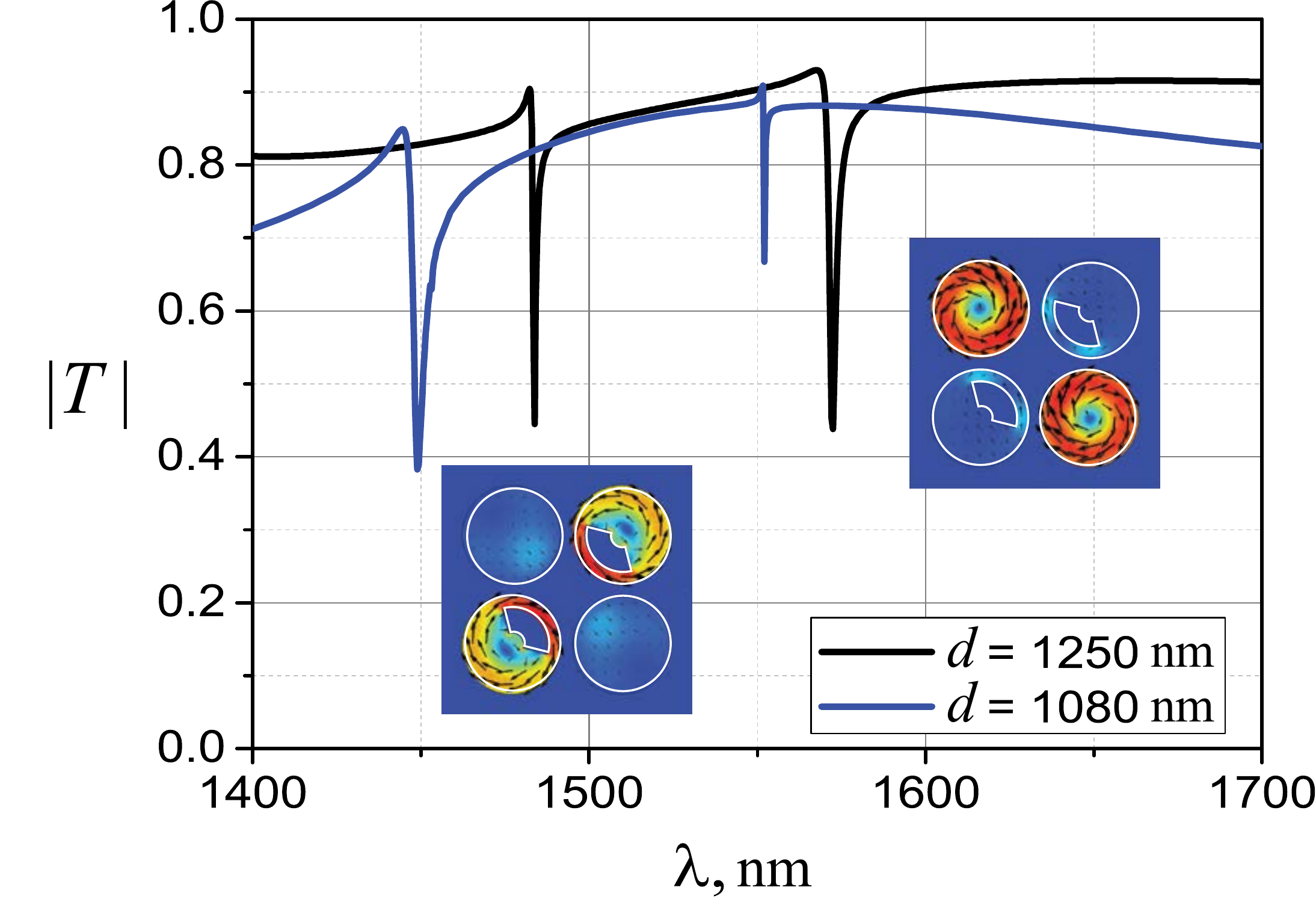}
\caption{Transmission coefficient of an all-dielectric metasurface with a diatomic design. A unit-cell cluster consists of a pair of two solid nanodisks and two nanodisks with coaxial-sector notches along diagonals. Insets show the displacement currents calculated inside the unit cell at the corresponding resonant wavelength; $d= 1250$~nm, $d_r=540$~nm, $h_d=120$~nm, $a_1=240$~nm, $a_h=a_1/2$, $w=a_1/2$, $\alpha=120$.}
\label{fig:symm_disks}
\end{figure}

In order to ensure certain generality of our analysis and conclusions, we have considered several other related metasurface structures composed on two types of resonators. It is expected that some other designs would allow to realize the toroidal response in the polarization-insensitive setting. An example of our numerical studies is shown in Fig.~\ref{fig:symm_disks}. We observe that only two closely spaced resonances appear in the wavelength range of interest, which correspond to the longitudinal MD excitation in the pairs of either solid nanodisks or nanodisks with notches. Thus, we confirm that the exitation of a strong toroidal responce in a polarization-insensitive all-dielectric metasurface is a highly nontrivial problem 
with no general answer.  

In summary, we have demonstrated that an all-dielectric metasurface composed of asymmetric clusters of  closely spaced dielectric nanodisks supports a strong axial toroidal moment. Such a toroidal moment originates from a trapped mode which is excited by an incident electromagnetic wave due to the symmetry breaking in the nanocluster. Our findings are supported by numerical simulations and qualitative studies, and they represent an important manifestation of the artificial magnetism that can be achieved in the all-dielectric metamaterials at optical frequencies.


\begin{acknowledgement}

The authors are indebted to the Australian Research Council. 

\end{acknowledgement}

\bibliography{toroidal_mode}

\providecommand{\latin}[1]{#1}
\makeatletter
\providecommand{\doi}
  {\begingroup\let\do\@makeother\dospecials
  \catcode`\{=1 \catcode`\}=2\doi@aux}
\providecommand{\doi@aux}[1]{\endgroup\texttt{#1}}
\makeatother
\providecommand*\mcitethebibliography{\thebibliography}
\csname @ifundefined\endcsname{endmcitethebibliography}
  {\let\endmcitethebibliography\endthebibliography}{}
\begin{mcitethebibliography}{49}
\providecommand*\natexlab[1]{#1}
\providecommand*\mciteSetBstSublistMode[1]{}
\providecommand*\mciteSetBstMaxWidthForm[2]{}
\providecommand*\mciteBstWouldAddEndPuncttrue
  {\def\EndOfBibitem{\unskip.}}
\providecommand*\mciteBstWouldAddEndPunctfalse
  {\let\EndOfBibitem\relax}
\providecommand*\mciteSetBstMidEndSepPunct[3]{}
\providecommand*\mciteSetBstSublistLabelBeginEnd[3]{}
\providecommand*\EndOfBibitem{}
\mciteSetBstSublistMode{f}
\mciteSetBstMaxWidthForm{subitem}{(\alph{mcitesubitemcount})}
\mciteSetBstSublistLabelBeginEnd
  {\mcitemaxwidthsubitemform\space}
  {\relax}
  {\relax}

\bibitem[Kaelberer \latin{et~al.}(2010)Kaelberer, Fedotov, Papasimakis, Tsai,
  and Zheludev]{Zheludev_Science_2010}
Kaelberer,~T.; Fedotov,~V.~A.; Papasimakis,~N.; Tsai,~D.~P.; Zheludev,~N.~I.
  \emph{Science} \textbf{2010}, \emph{330}, 1510--1512\relax
\mciteBstWouldAddEndPuncttrue
\mciteSetBstMidEndSepPunct{\mcitedefaultmidpunct}
{\mcitedefaultendpunct}{\mcitedefaultseppunct}\relax
\EndOfBibitem
\bibitem[Dubovik \latin{et~al.}(1986)Dubovik, Tosunyan, and
  Tugushev]{Dubovik_JETP_1986}
Dubovik,~V.~M.; Tosunyan,~L.~A.; Tugushev,~V.~V. \emph{Zh. Eksp. Teor. Fiz.}
  \textbf{1986}, \emph{90}, 590--605\relax
\mciteBstWouldAddEndPuncttrue
\mciteSetBstMidEndSepPunct{\mcitedefaultmidpunct}
{\mcitedefaultendpunct}{\mcitedefaultseppunct}\relax
\EndOfBibitem
\bibitem[Thorner \latin{et~al.}(2014)Thorner, Kiat, Bogicevic, and
  Kornev]{Thorner_PhysRevB_2014}
Thorner,~G.; Kiat,~J.-M.; Bogicevic,~C.; Kornev,~I. \emph{Phys. Rev. B}
  \textbf{2014}, \emph{89}, 220103\relax
\mciteBstWouldAddEndPuncttrue
\mciteSetBstMidEndSepPunct{\mcitedefaultmidpunct}
{\mcitedefaultendpunct}{\mcitedefaultseppunct}\relax
\EndOfBibitem
\bibitem[Marinov \latin{et~al.}(2007)Marinov, Boardman, Fedotov, and
  Zheludev]{Zheludev_NewJPhys_2007}
Marinov,~K.; Boardman,~A.~D.; Fedotov,~V.~A.; Zheludev,~N. \emph{New J. Phys.}
  \textbf{2007}, \emph{9}, 324\relax
\mciteBstWouldAddEndPuncttrue
\mciteSetBstMidEndSepPunct{\mcitedefaultmidpunct}
{\mcitedefaultendpunct}{\mcitedefaultseppunct}\relax
\EndOfBibitem
\bibitem[Savinov \latin{et~al.}(2014)Savinov, Fedotov, and
  Zheludev]{Savinov_PhysRevB_2014}
Savinov,~V.; Fedotov,~V.~A.; Zheludev,~N.~I. \emph{Phys. Rev. B} \textbf{2014},
  \emph{89}, 205112\relax
\mciteBstWouldAddEndPuncttrue
\mciteSetBstMidEndSepPunct{\mcitedefaultmidpunct}
{\mcitedefaultendpunct}{\mcitedefaultseppunct}\relax
\EndOfBibitem
\bibitem[Naumov \latin{et~al.}(2004)Naumov, Bellaiche, and
  Fu]{Naumov_Nature_2004}
Naumov,~I.~I.; Bellaiche,~L.; Fu,~H. \emph{Nature} \textbf{2004}, \emph{432},
  737--740\relax
\mciteBstWouldAddEndPuncttrue
\mciteSetBstMidEndSepPunct{\mcitedefaultmidpunct}
{\mcitedefaultendpunct}{\mcitedefaultseppunct}\relax
\EndOfBibitem
\bibitem[Spaldin \latin{et~al.}(2008)Spaldin, Fiebig, and
  Mostovoy]{Spaldin_JPhysCondensMatter_2008}
Spaldin,~N.~A.; Fiebig,~M.; Mostovoy,~M. \emph{J. Phys. Condens. Matter}
  \textbf{2008}, \emph{20}, 434203\relax
\mciteBstWouldAddEndPuncttrue
\mciteSetBstMidEndSepPunct{\mcitedefaultmidpunct}
{\mcitedefaultendpunct}{\mcitedefaultseppunct}\relax
\EndOfBibitem
\bibitem[Ceulemans \latin{et~al.}(1998)Ceulemans, Chibotaru, and
  Fowler]{Ceulemans_PhysRevLett_1998}
Ceulemans,~A.; Chibotaru,~L.~F.; Fowler,~P.~W. \emph{Phys. Rev. Lett.}
  \textbf{1998}, \emph{80}, 1861--1864\relax
\mciteBstWouldAddEndPuncttrue
\mciteSetBstMidEndSepPunct{\mcitedefaultmidpunct}
{\mcitedefaultendpunct}{\mcitedefaultseppunct}\relax
\EndOfBibitem
\bibitem[Yamaguchi and Kimura(2013)Yamaguchi, and
  Kimura]{Yamaguchi_NatureCommun_2013}
Yamaguchi,~Y.; Kimura,~T. \emph{Nat. Commun.} \textbf{2013}, \emph{4},
  2063\relax
\mciteBstWouldAddEndPuncttrue
\mciteSetBstMidEndSepPunct{\mcitedefaultmidpunct}
{\mcitedefaultendpunct}{\mcitedefaultseppunct}\relax
\EndOfBibitem
\bibitem[Talebi \latin{et~al.}(2017)Talebi, Guo, and {van
  Aken}]{Talebi_nanoph_2017}
Talebi,~N.; Guo,~S.; {van Aken},~P. \emph{Nanophotonics} \textbf{2017},
  \emph{7}, 93--110\relax
\mciteBstWouldAddEndPuncttrue
\mciteSetBstMidEndSepPunct{\mcitedefaultmidpunct}
{\mcitedefaultendpunct}{\mcitedefaultseppunct}\relax
\EndOfBibitem
\bibitem[Zheludev and Kivshar(2012)Zheludev, and
  Kivshar]{Zheludev_NatMater_2012}
Zheludev,~N.~I.; Kivshar,~Y.~S. \emph{Nat. Mater.} \textbf{2012}, \emph{11},
  917--924\relax
\mciteBstWouldAddEndPuncttrue
\mciteSetBstMidEndSepPunct{\mcitedefaultmidpunct}
{\mcitedefaultendpunct}{\mcitedefaultseppunct}\relax
\EndOfBibitem
\bibitem[Zheludev(2015)]{Zheludev_Science_2015}
Zheludev,~N.~I. \emph{Science} \textbf{2015}, \emph{348}, 973--974\relax
\mciteBstWouldAddEndPuncttrue
\mciteSetBstMidEndSepPunct{\mcitedefaultmidpunct}
{\mcitedefaultendpunct}{\mcitedefaultseppunct}\relax
\EndOfBibitem
\bibitem[Papasimakis \latin{et~al.}(2009)Papasimakis, Fedotov, Marinov, and
  Zheludev]{Papasimakis_PhysRevLett_2009}
Papasimakis,~N.; Fedotov,~V.~A.; Marinov,~K.; Zheludev,~N.~I. \emph{Phys. Rev.
  Lett.} \textbf{2009}, \emph{103}, 093901\relax
\mciteBstWouldAddEndPuncttrue
\mciteSetBstMidEndSepPunct{\mcitedefaultmidpunct}
{\mcitedefaultendpunct}{\mcitedefaultseppunct}\relax
\EndOfBibitem
\bibitem[Dong \latin{et~al.}(2012)Dong, Ni, Zhu, Yin, and
  Zhang]{Dong_OptExpress_2012}
Dong,~Z.-G.; Ni,~P.; Zhu,~J.; Yin,~X.; Zhang,~X. \emph{Opt. Express}
  \textbf{2012}, \emph{20}, 13065--13070\relax
\mciteBstWouldAddEndPuncttrue
\mciteSetBstMidEndSepPunct{\mcitedefaultmidpunct}
{\mcitedefaultendpunct}{\mcitedefaultseppunct}\relax
\EndOfBibitem
\bibitem[Guo \latin{et~al.}(2012)Guo, Li, Ye, Xiao, and
  Yang]{Guo_EurPhysJB_2012}
Guo,~L.~Y.; Li,~M.~H.; Ye,~Q.~W.; Xiao,~B.~X.; Yang,~H.~L. \emph{Eur. Phys. J.
  B} \textbf{2012}, \emph{85}, 208\relax
\mciteBstWouldAddEndPuncttrue
\mciteSetBstMidEndSepPunct{\mcitedefaultmidpunct}
{\mcitedefaultendpunct}{\mcitedefaultseppunct}\relax
\EndOfBibitem
\bibitem[Fan \latin{et~al.}(2013)Fan, Wei, Li, Chen, and
  Soukoulis]{Fan_PhysRevB_2013}
Fan,~Y.; Wei,~Z.; Li,~H.; Chen,~H.; Soukoulis,~C.~M. \emph{Phys. Rev. B}
  \textbf{2013}, \emph{87}, 115417\relax
\mciteBstWouldAddEndPuncttrue
\mciteSetBstMidEndSepPunct{\mcitedefaultmidpunct}
{\mcitedefaultendpunct}{\mcitedefaultseppunct}\relax
\EndOfBibitem
\bibitem[Raybould \latin{et~al.}(2016)Raybould, Fedotov, Papasimakis, Kuprov,
  Youngs, Chen, Tsai, and Zheludev]{zheludev_PhysRevB_2016}
Raybould,~T.~A.; Fedotov,~V.~A.; Papasimakis,~N.; Kuprov,~I.; Youngs,~I.~J.;
  Chen,~W.~T.; Tsai,~D.~P.; Zheludev,~N.~I. \emph{Phys. Rev. B} \textbf{2016},
  \emph{94}, 035119\relax
\mciteBstWouldAddEndPuncttrue
\mciteSetBstMidEndSepPunct{\mcitedefaultmidpunct}
{\mcitedefaultendpunct}{\mcitedefaultseppunct}\relax
\EndOfBibitem
\bibitem[Stenishchev and Basharin(2017)Stenishchev, and
  Basharin]{Stenishchev_SciRep_2017}
Stenishchev,~I.~V.; Basharin,~A.~A. \emph{Sci. Rep.} \textbf{2017}, \emph{7},
  9468\relax
\mciteBstWouldAddEndPuncttrue
\mciteSetBstMidEndSepPunct{\mcitedefaultmidpunct}
{\mcitedefaultendpunct}{\mcitedefaultseppunct}\relax
\EndOfBibitem
\bibitem[Dong \latin{et~al.}(2012)Dong, Zhu, Rho, Li, Lu, Yin, and
  Zhang]{Dong_ApplPhysLett_2012}
Dong,~Z.-G.; Zhu,~J.; Rho,~J.; Li,~J.-Q.; Lu,~C.; Yin,~X.; Zhang,~X.
  \emph{Appl. Phys. Lett.} \textbf{2012}, \emph{101}, 144105\relax
\mciteBstWouldAddEndPuncttrue
\mciteSetBstMidEndSepPunct{\mcitedefaultmidpunct}
{\mcitedefaultendpunct}{\mcitedefaultseppunct}\relax
\EndOfBibitem
\bibitem[Dong \latin{et~al.}(2013)Dong, Zhu, Yin, Li, Lu, and
  Zhang]{Dong_PhysRevB_2013}
Dong,~Z.-G.; Zhu,~J.; Yin,~X.; Li,~J.; Lu,~C.; Zhang,~X. \emph{Phys. Rev. B}
  \textbf{2013}, \emph{87}, 245429\relax
\mciteBstWouldAddEndPuncttrue
\mciteSetBstMidEndSepPunct{\mcitedefaultmidpunct}
{\mcitedefaultendpunct}{\mcitedefaultseppunct}\relax
\EndOfBibitem
\bibitem[Gupta \latin{et~al.}(2016)Gupta, Savinov, Xu, Cong, Dayal, Wang,
  Zhang, Zheludev, and Singh]{Gupta_ADMA_2016}
Gupta,~M.; Savinov,~V.; Xu,~N.; Cong,~L.; Dayal,~G.; Wang,~S.; Zhang,~W.;
  Zheludev,~N.~I.; Singh,~R. \emph{Adv. Materials} \textbf{2016}, \emph{28},
  8206--8211\relax
\mciteBstWouldAddEndPuncttrue
\mciteSetBstMidEndSepPunct{\mcitedefaultmidpunct}
{\mcitedefaultendpunct}{\mcitedefaultseppunct}\relax
\EndOfBibitem
\bibitem[\"Og\"ut \latin{et~al.}(2012)\"Og\"ut, Talebi, Vogelgesang, Sigle, and
  van Aken]{Ogut_NanoLett_2012}
\"Og\"ut,~B.; Talebi,~N.; Vogelgesang,~R.; Sigle,~W.; van Aken,~P.~A.
  \emph{Nano Lett.} \textbf{2012}, \emph{12}, 5239--5244\relax
\mciteBstWouldAddEndPuncttrue
\mciteSetBstMidEndSepPunct{\mcitedefaultmidpunct}
{\mcitedefaultendpunct}{\mcitedefaultseppunct}\relax
\EndOfBibitem
\bibitem[Huang \latin{et~al.}(2012)Huang, Chen, Wu, Fedotov, Savinov, Ho, Chau,
  Zheludev, and Tsai]{Huang_OptExpress_2012}
Huang,~Y.-W.; Chen,~W.~T.; Wu,~P.~C.; Fedotov,~V.; Savinov,~V.; Ho,~Y.~Z.;
  Chau,~Y.-F.; Zheludev,~N.~I.; Tsai,~D.~P. \emph{Opt. Express} \textbf{2012},
  \emph{20}, 1760--1768\relax
\mciteBstWouldAddEndPuncttrue
\mciteSetBstMidEndSepPunct{\mcitedefaultmidpunct}
{\mcitedefaultendpunct}{\mcitedefaultseppunct}\relax
\EndOfBibitem
\bibitem[Huang \latin{et~al.}(2013)Huang, Wang, Lu, Zhang, Min, Lin, Ti, Xu,
  He, Yue, and Zhu]{Huang_SciRep_2013}
Huang,~F.; Wang,~Z.; Lu,~X.; Zhang,~J.; Min,~K.; Lin,~W.; Ti,~R.; Xu,~T.;
  He,~J.; Yue,~C.; Zhu,~J. \emph{Sci. Rep.} \textbf{2013}, \emph{3}, 2907\relax
\mciteBstWouldAddEndPuncttrue
\mciteSetBstMidEndSepPunct{\mcitedefaultmidpunct}
{\mcitedefaultendpunct}{\mcitedefaultseppunct}\relax
\EndOfBibitem
\bibitem[Decker \latin{et~al.}(2015)Decker, Staude, Falkner, Dominguez, Neshev,
  Brener, Pertsch, and Kivshar]{Decker_AdvOptMat_2015}
Decker,~M.; Staude,~I.; Falkner,~M.; Dominguez,~J.; Neshev,~D.~N.; Brener,~I.;
  Pertsch,~T.; Kivshar,~Y.~S. \emph{Adv. Opt. Mater.} \textbf{2015}, \emph{3},
  813–--820\relax
\mciteBstWouldAddEndPuncttrue
\mciteSetBstMidEndSepPunct{\mcitedefaultmidpunct}
{\mcitedefaultendpunct}{\mcitedefaultseppunct}\relax
\EndOfBibitem
\bibitem[Jahani and Jacob(2016)Jahani, and Jacob]{jahani_NatNano_2016}
Jahani,~S.; Jacob,~Z. \emph{Nat. Nanotechnol.} \textbf{2016}, \emph{11},
  23--36\relax
\mciteBstWouldAddEndPuncttrue
\mciteSetBstMidEndSepPunct{\mcitedefaultmidpunct}
{\mcitedefaultendpunct}{\mcitedefaultseppunct}\relax
\EndOfBibitem
\bibitem[Kuznetsov \latin{et~al.}(2016)Kuznetsov, Miroshnichenko, Brongersma,
  Kivshar, and Luk'yanchuk]{Kuznetsov_Science_2016}
Kuznetsov,~A.~I.; Miroshnichenko,~A.~E.; Brongersma,~M.~L.; Kivshar,~Y.~S.;
  Luk'yanchuk,~B. \emph{Science} \textbf{2016}, \emph{354}\relax
\mciteBstWouldAddEndPuncttrue
\mciteSetBstMidEndSepPunct{\mcitedefaultmidpunct}
{\mcitedefaultendpunct}{\mcitedefaultseppunct}\relax
\EndOfBibitem
\bibitem[Kruk and Kivshar(2017)Kruk, and Kivshar]{Kruk_ACSPhotonics_2017}
Kruk,~S.; Kivshar,~Y. \emph{ACS Photonics} \textbf{2017}, \emph{4},
  2638--2649\relax
\mciteBstWouldAddEndPuncttrue
\mciteSetBstMidEndSepPunct{\mcitedefaultmidpunct}
{\mcitedefaultendpunct}{\mcitedefaultseppunct}\relax
\EndOfBibitem
\bibitem[Zhao \latin{et~al.}(2009)Zhao, Zhou, Zhang, and
  Lippens]{Zhao_MatToday_2009}
Zhao,~Q.; Zhou,~J.; Zhang,~F.; Lippens,~D. \emph{Mater. Today} \textbf{2009},
  \emph{12}, 60--69\relax
\mciteBstWouldAddEndPuncttrue
\mciteSetBstMidEndSepPunct{\mcitedefaultmidpunct}
{\mcitedefaultendpunct}{\mcitedefaultseppunct}\relax
\EndOfBibitem
\bibitem[Miroshnichenko \latin{et~al.}(2015)Miroshnichenko, Evlyukhin, Yu,
  Bakker, Chipouline, Kuznetsov, Luk'yanchuk, Chichkov, and
  Kivshar]{Miroshnichenko_NatCommun_2017}
Miroshnichenko,~A.~E.; Evlyukhin,~A.~B.; Yu,~Y.~F.; Bakker,~R.~M.;
  Chipouline,~A.; Kuznetsov,~A.~I.; Luk'yanchuk,~B.; Chichkov,~B.~N.;
  Kivshar,~Y.~S. \emph{Nat. Commun.} \textbf{2015}, \emph{6}, 8069\relax
\mciteBstWouldAddEndPuncttrue
\mciteSetBstMidEndSepPunct{\mcitedefaultmidpunct}
{\mcitedefaultendpunct}{\mcitedefaultseppunct}\relax
\EndOfBibitem
\bibitem[Basharin \latin{et~al.}(2015)Basharin, Kafesaki, Economou, Soukoulis,
  Fedotov, Savinov, and Zheludev]{Basharin_PhysRevX_2015}
Basharin,~A.~A.; Kafesaki,~M.; Economou,~E.~N.; Soukoulis,~C.~M.;
  Fedotov,~V.~A.; Savinov,~V.; Zheludev,~N.~I. \emph{Phys. Rev. X}
  \textbf{2015}, \emph{5}, 011036\relax
\mciteBstWouldAddEndPuncttrue
\mciteSetBstMidEndSepPunct{\mcitedefaultmidpunct}
{\mcitedefaultendpunct}{\mcitedefaultseppunct}\relax
\EndOfBibitem
\bibitem[Tasolamprou \latin{et~al.}(2016)Tasolamprou, Tsilipakos, Kafesaki,
  Soukoulis, and Economou]{Soukoulis_PhysRevB_2016}
Tasolamprou,~A.~C.; Tsilipakos,~O.; Kafesaki,~M.; Soukoulis,~C.~M.;
  Economou,~E.~N. \emph{Phys. Rev. B} \textbf{2016}, \emph{94}, 205433\relax
\mciteBstWouldAddEndPuncttrue
\mciteSetBstMidEndSepPunct{\mcitedefaultmidpunct}
{\mcitedefaultendpunct}{\mcitedefaultseppunct}\relax
\EndOfBibitem
\bibitem[Bao \latin{et~al.}(2015)Bao, Zhu, and Fang]{Bao_SciRep_2015}
Bao,~Y.; Zhu,~X.; Fang,~Z. \emph{Sci. Rep.} \textbf{2015}, \emph{5},
  11793\relax
\mciteBstWouldAddEndPuncttrue
\mciteSetBstMidEndSepPunct{\mcitedefaultmidpunct}
{\mcitedefaultendpunct}{\mcitedefaultseppunct}\relax
\EndOfBibitem
\bibitem[Guo \latin{et~al.}(2014)Guo, Li, Huang, and
  Yang]{Gui_ApplPhysLett_2014}
Guo,~L.-Y.; Li,~M.-H.; Huang,~X.-J.; Yang,~H.-L. \emph{Appl. Phys. Lett.}
  \textbf{2014}, \emph{105}, 033507\relax
\mciteBstWouldAddEndPuncttrue
\mciteSetBstMidEndSepPunct{\mcitedefaultmidpunct}
{\mcitedefaultendpunct}{\mcitedefaultseppunct}\relax
\EndOfBibitem
\bibitem[Prosvirnin and Zouhdi(2003)Prosvirnin, and
  Zouhdi]{Zouhdi_Advances_2003}
Prosvirnin,~S.; Zouhdi,~S. Resonances of closed modes in thin arrays of complex
  particles. Advances in Electromagnetics of Complex Media and Metamaterials.
  Printed in the Netherlands, 2003; pp 281--290\relax
\mciteBstWouldAddEndPuncttrue
\mciteSetBstMidEndSepPunct{\mcitedefaultmidpunct}
{\mcitedefaultendpunct}{\mcitedefaultseppunct}\relax
\EndOfBibitem
\bibitem[Fedotov \latin{et~al.}(2007)Fedotov, Rose, Prosvirnin, Papasimakis,
  and Zheludev]{Fedotov_PhysRevLett_2007}
Fedotov,~V.~A.; Rose,~M.; Prosvirnin,~S.~L.; Papasimakis,~N.; Zheludev,~N.~I.
  \emph{Phys. Rev. Lett.} \textbf{2007}, \emph{99}, 147401\relax
\mciteBstWouldAddEndPuncttrue
\mciteSetBstMidEndSepPunct{\mcitedefaultmidpunct}
{\mcitedefaultendpunct}{\mcitedefaultseppunct}\relax
\EndOfBibitem
\bibitem[Khardikov \latin{et~al.}(2012)Khardikov, Iarko, and
  Prosvirnin]{Khardikov_JOpt_2012}
Khardikov,~V.~V.; Iarko,~E.~O.; Prosvirnin,~S.~L. \emph{J. Opt.} \textbf{2012},
  \emph{14}, 035103\relax
\mciteBstWouldAddEndPuncttrue
\mciteSetBstMidEndSepPunct{\mcitedefaultmidpunct}
{\mcitedefaultendpunct}{\mcitedefaultseppunct}\relax
\EndOfBibitem
\bibitem[Khardikov \latin{et~al.}(2016)Khardikov, Mladyonov, Prosvirnin, and
  Tuz]{Khardikov2016}
Khardikov,~V.; Mladyonov,~P.; Prosvirnin,~S.; Tuz,~V. In \emph{Contemporary
  Optoelectronics: Materials, Metamaterials and Device Applications};
  Shulika,~O., Sukhoivanov,~I., Eds.; Springer Netherlands: Dordrecht, 2016;
  Chapter 5, pp 81--98\relax
\mciteBstWouldAddEndPuncttrue
\mciteSetBstMidEndSepPunct{\mcitedefaultmidpunct}
{\mcitedefaultendpunct}{\mcitedefaultseppunct}\relax
\EndOfBibitem
\bibitem[com()]{comsol}
Comsol Application Gallery ID: 15711.
  \url{https://www.comsol.com/model/frequency-selective-surface-periodic-complementary-split-ring-resonator-15711}\relax
\mciteBstWouldAddEndPuncttrue
\mciteSetBstMidEndSepPunct{\mcitedefaultmidpunct}
{\mcitedefaultendpunct}{\mcitedefaultseppunct}\relax
\EndOfBibitem
\bibitem[Munk(2000)]{munk_2000}
Munk,~B.~A. \emph{Frequency selective surfaces: Theory and design}; Wiley, New
  York, 2000\relax
\mciteBstWouldAddEndPuncttrue
\mciteSetBstMidEndSepPunct{\mcitedefaultmidpunct}
{\mcitedefaultendpunct}{\mcitedefaultseppunct}\relax
\EndOfBibitem
\bibitem[Snitzer(1961)]{Snitzer_JOptAm_61}
Snitzer,~E. \emph{J. Opt. Soc. Am.} \textbf{1961}, \emph{51}, 491--498\relax
\mciteBstWouldAddEndPuncttrue
\mciteSetBstMidEndSepPunct{\mcitedefaultmidpunct}
{\mcitedefaultendpunct}{\mcitedefaultseppunct}\relax
\EndOfBibitem
\bibitem[Marcatili and Schmeltzer(1964)Marcatili, and
  Schmeltzer]{marcatili1964hollow}
Marcatili,~E. A.~J.; Schmeltzer,~R.~A. \emph{Bell Labs Tech. J.} \textbf{1964},
  \emph{43}, 1783--1809\relax
\mciteBstWouldAddEndPuncttrue
\mciteSetBstMidEndSepPunct{\mcitedefaultmidpunct}
{\mcitedefaultendpunct}{\mcitedefaultseppunct}\relax
\EndOfBibitem
\bibitem[Limonov \latin{et~al.}(2017)Limonov, Rybin, Poddubny, and
  Kivshar]{Kivshar_NatPhot_2017}
Limonov,~M.~F.; Rybin,~M.~V.; Poddubny,~A.~N.; Kivshar,~Y.~S. \emph{Nat.
  Photon.} \textbf{2017}, \emph{11}, 543--554\relax
\mciteBstWouldAddEndPuncttrue
\mciteSetBstMidEndSepPunct{\mcitedefaultmidpunct}
{\mcitedefaultendpunct}{\mcitedefaultseppunct}\relax
\EndOfBibitem
\bibitem[Evans \latin{et~al.}(1993)Evans, Linton, and
  Ursell]{Evans_MechApplMath_1993}
Evans,~D.~V.; Linton,~C.~M.; Ursell,~F. \emph{Q. J. Mech. Appl. Math.}
  \textbf{1993}, \emph{46}, 253--274\relax
\mciteBstWouldAddEndPuncttrue
\mciteSetBstMidEndSepPunct{\mcitedefaultmidpunct}
{\mcitedefaultendpunct}{\mcitedefaultseppunct}\relax
\EndOfBibitem
\bibitem[Evans \latin{et~al.}(1994)Evans, Levitin, and
  Vassiliev]{Evans_JFluidMech_1994}
Evans,~D.~V.; Levitin,~M.; Vassiliev,~D. \emph{J. Fluid Mech.} \textbf{1994},
  \emph{261}, 21--31\relax
\mciteBstWouldAddEndPuncttrue
\mciteSetBstMidEndSepPunct{\mcitedefaultmidpunct}
{\mcitedefaultendpunct}{\mcitedefaultseppunct}\relax
\EndOfBibitem
\bibitem[Tuz \latin{et~al.}(2010)Tuz, Prosvirnin, and
  Kochetova]{tuz_PhysRevB_2010}
Tuz,~V.~R.; Prosvirnin,~S.~L.; Kochetova,~L.~A. \emph{Phys. Rev. B}
  \textbf{2010}, \emph{82}, 233402\relax
\mciteBstWouldAddEndPuncttrue
\mciteSetBstMidEndSepPunct{\mcitedefaultmidpunct}
{\mcitedefaultendpunct}{\mcitedefaultseppunct}\relax
\EndOfBibitem
\bibitem[Tuz and Prosvirnin(2011)Tuz, and
  Prosvirnin]{tuz_EurPhysJApplPhys_2011}
Tuz,~V.; Prosvirnin,~S. \emph{Eur. Phys. J. App. Phys.} \textbf{2011},
  \emph{56}, 30401\relax
\mciteBstWouldAddEndPuncttrue
\mciteSetBstMidEndSepPunct{\mcitedefaultmidpunct}
{\mcitedefaultendpunct}{\mcitedefaultseppunct}\relax
\EndOfBibitem
\bibitem[Tuz \latin{et~al.}(2014)Tuz, Novitsky, Mladyonov, Prosvirnin, and
  Novitsky]{Tuz_JOSAB_2014}
Tuz,~V.~R.; Novitsky,~D.~V.; Mladyonov,~P.~L.; Prosvirnin,~S.~L.;
  Novitsky,~A.~V. \emph{J. Opt. Soc. Am. B} \textbf{2014}, \emph{31},
  2095--2103\relax
\mciteBstWouldAddEndPuncttrue
\mciteSetBstMidEndSepPunct{\mcitedefaultmidpunct}
{\mcitedefaultendpunct}{\mcitedefaultseppunct}\relax
\EndOfBibitem
\end{mcitethebibliography}
\newpage
\section*{For Table of Contents Use Only}

\subsection*{Manuscript title:} All-dielectric resonant metasurfaces with a~strong toroidal response

\subsection*{Names of authors:} Vladimir R. Tuz, Vyacheslav V. Khardikov, Yuri S. Kivshar

\subsection*{Brief synopsis:} We demonstrate how to create all-dielectric metasurfaces with a strong toroidal response by arranging two types of nanodisks into asymmetric quadrumer clusters. We demonstrate that a strong axial toroidal response of the metasurface is related to conditions of the trapped (dark) mode that is excited due the symmetry breaking in the cluster. We study the correlation between the toroidal response and asymmetry in the metasurface and nanocluster geometries, which appears from the different diameters of nanodisks or notches introduced into the nanodisks.

\begin{figure}[ht!]
\centering\includegraphics[width=6cm]{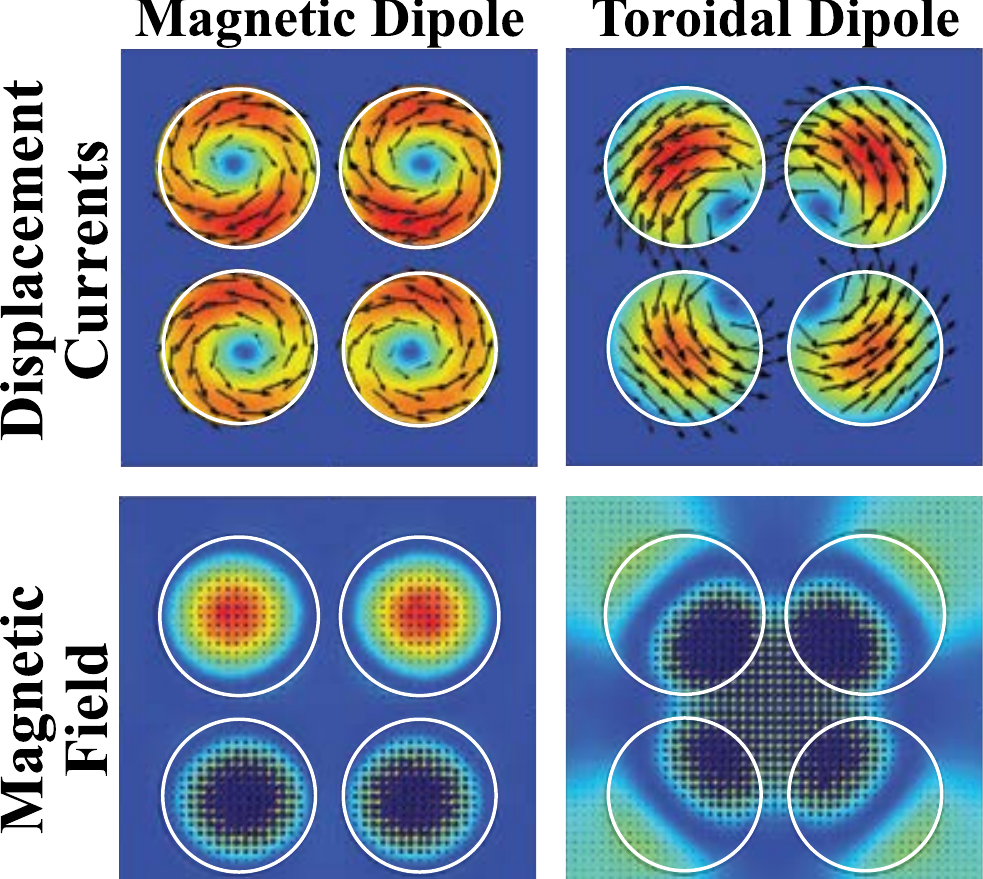}
\caption{ToC-image}
\label{fig:toc}

\end{figure}
\end{document}